\documentclass[AMA,STIX1COL]{WileyNJD-v2}
\usepackage{moreverb}

\newcommand\BibTeX{{\rmfamily B\kern-.05em \textsc{i\kern-.025em b}\kern-.08em
T\kern-.1667em\lower.7ex\hbox{E}\kern-.125emX}}

\articletype{Article Type}%

\received{<day> <Month>, <year>}
\revised{<day> <Month>, <year>}
\accepted{<day> <Month>, <year>}

\begin{document}

\title{Thermodynamic and transport properties of plasmas: low-density benchmarks}

\author{G. R\"opke} 

\address{Institut f\"ur Physik, Universit\"at Rostock, 18051 Rostock, Germany}

\corres{ \email{gerd.roepke@uni-rostock.de}}

\abstract[Abstract]{Physical properties of plasmas such as equations of state and transport coefficients are expressed in terms of correlation functions, which can be calculated using various approaches (analytical theory, numerical simulations). 
The method of Green's functions provides benchmark values for these properties in the low-density limit. 
For the equation of state and electrical conductivity, expansions with respect to density (virial expansions) are considered. 
Comparison of analytical results with numerical simulations is used to verify theory, to prove the accuracy of simulations, and to establish interpolation formulas.}


\maketitle

\section{Plasma properties and correlation functions}\label{sec0}

Plasmas consist of charged particles, number $N_i$ of species $i$ in the volume $\Omega$, which interact via the Coulomb law. If we denote the charge of the component $i$ by $Z_ie$, we obtain ($\epsilon_0$ is the permittivity of the vacuum)
\begin{equation}
V^{\rm Coul}_{ij}(r) = \frac{Z_iZ_j e^2}{4 \pi \epsilon_0 r}.
\end{equation}
In general, an additional short-range interaction may occur. Examples are the homogeneous electron gas (uniform electron gas UEG), where the electrons move over a positively charged background to realize charge neutrality, or the two-component Hydrogen plasma, consisting of electrons and protons, where the particle density is $n_e=n_p$ to maintain charge neutrality. In thermodynamic equilibrium, the state of the plasma is determined by the temperature $T$ in addition to the densities $n_i=N_i/\Omega$ of the components or the corresponding chemical potentials $\mu_i$. The relationships between the various state variables such as internal energy $U$, free energy $F$, entropy $S$, pressure $P$, etc. are called equations of state (EoS). All thermodynamic properties can be derived from a thermodynamic potential, for example is $F(\Omega, N_i,T)$ as function of $\Omega, N_i,T$ a thermodynamic potential.

Statistical physics allows to calculate the thermodynamic properties from the microscopic properties, i.e. from the Hamiltonian $H=H_{\rm kin} + V$, with kinetic energy $H_{\rm kin}=\sum_i \sum_k^{N_i} p_{i,k}^2/2 m_i$ and potential energy $V=(1/2) \sum_{i,k \neq j,l} V({\bf r}_{i,k}-{\bf r}_{j,l})$. 
To calculate physical quantities, various expressions can be used. For example, for classical systems we can start from the well-known partition function $Z_{\rm can}(\Omega, N_i,T)$ with $F(\Omega, N_i,T)=-k_BT \ln Z_{\rm can}(\Omega, N_i,T)$.
For quantum systems, it is convenient to work with the grand  canonical ensemble defined by $\beta=1/k_BT$ and the chemical potentials $\mu_i$.
Then the single-particle distribution functions for the ideal quantum system ($V=0$) have a simple form, the Fermi or Bose distribution. 
In second quantization, we introduce $a^+_{i,k},a_{i,k}$ as a creation or annihilation operator for particles of species $i$ in the quantum state $k=\{\hbar {\bf k}, \sigma\}$, which denotes momentum vector and spin. 
The occupation number of this quantum state is given as
\begin{equation}
\label{fik}
f_{i,k} = \langle a^+_{i,k}a_{i,k} \rangle = \frac{1}{Z_{\rm gr.can}}
{\rm Tr} \left\{e^{-\beta (H-\sum_i \mu_i N_i)}a^+_{i,k}a_{i,k}\right\},\qquad Z_{\rm gr.can}= {\rm Tr} e^{-\beta (H-\sum_i \mu_i N_i)}
\end{equation}
with $H$ as the Hamiltonian in second quantization and $N_i=\sum_k a^+_{i,k}a_{i,k}$. The relation between the densities and the chemical potentials is given as follows.
\begin{equation}
\label{simpleEOS}
n_i(T,\{\mu_j\})=\frac{1}{\Omega}\langle N_i \rangle =\frac{1}{\Omega}\sum_k f_{i,k}.
\end{equation}

It is convenient to introduce a $\tau$-dependent correlation function as a generalization of (\ref{fik})  that contains dynamic information,
\begin{equation}
\label{specdens}
 \langle a^+_{j,l}e^{\tau (H-\sum_i \mu_i N_i)} a_{i,k}e^{-\tau (H-\sum_i \mu_i N_i)} \rangle =\int_{-\infty}^\infty \frac{d \omega}{2 \pi}e^{-\omega \tau}I_{ik,jl}(\omega).
\end{equation}
The spectral density $I_{ik,jl}(\omega)$ is related to the spectral function $A_{ik,jl}(\omega)=(1 + e^{\beta \omega})I_{ik,jl}(\omega)$ (Fermi statistics).
An exact expression for the EoS is found if the spectral function is known,
\begin{equation}
\label{EOS}
n_i(T,\{\mu_j\})=\frac{1}{\Omega}\sum_k \int_{-\infty}^\infty \frac{d \omega}{2 \pi}\frac{1}{e^{\beta \omega}+1} A_{ik,ik}(\omega).
\end{equation}
The spectral function which is diagonal with respect to $\{i,k\}$ for a homogeneous system,
is related to the self-energy $\Sigma_{i,k}(z)$ for which a systematic evaluation applying diagram techniques is possible, see \cite{FW,GRBuch}:
\begin{equation}
\label{spectral}
 A_{ik}(\omega) = \frac{2 {\rm Im}\Sigma_{i,k}(\omega-i0)}{[\omega - \epsilon_{i,k}- {\rm Re} \Sigma_{i,k}(\omega)]^2 + 
[{\rm Im}\Sigma_{i,k}(\omega-i0)]^2 }\,,
\end{equation}
$ \epsilon_{i,k}=\hbar^2k^2/2m_i-\mu_i$ is the kinetic energy shifted by the chemical potential.

The electrical conductivity $\sigma(T,n)$ of low-density plasmas was first calculated in the framework of kinetic theory. 
In a seminal work~\cite{Spitzer53}, Spitzer and H{\"a}rm determined $\sigma$ of the fully ionized Hydrogen plasma by solving a Fokker-Planck equation.
To calculate $\sigma (T,n)$ in a wide range of temperature $T$ and particle density $n$, a quantum statistical many-particle theory is needed that describes screening, correlations, and degeneracy effects in a systematic way. 
A generalized linear response theory~\cite{Roep88,RR89,Redmer97} has been elaborated that expresses transport coefficients in terms of equilibrium correlation functions (fluctuation-dissipation theorems).

An example is the Kubo formula~\cite{Kubo66} which relates the transport coefficient $\sigma$ to  the electron current-current correlation function, 
\begin{equation}
\sigma(T,n)=\frac{e^2}{m^2_e k_BT \Omega}\langle P;P \rangle_{i \epsilon}
\label{Kubo}
\end{equation}
with the total momentum of the electrons $P=\sum_k \hbar k_x a^+_{e,k} a_{e,k}$ in $x$ direction (the small ion contribution to the electrical current may be added). 
The thermodynamic correlation function is the Laplace transform of the Kubo scalar product (the particle number is assumed to commute with the observables),
\begin{equation}
\label{Laplace}
\langle A;B \rangle_z=
\int\limits_0^{\infty}d t\, e^{i zt}\frac{1}{\beta}\int\limits_0^\beta d \tau \langle e^{(i/\hbar)(t-i\hbar \tau)H} A e^{-(i/\hbar)(t-i\hbar \tau)H} B\rangle\,.
\end{equation}
For more details on generalized linear response theory and the evaluation of correlation functions using the method of thermodynamic Green's functions, see \cite{GRBuch}.
For the relationship between generalized linear response theory and kinetic theory, see~\cite{Reinholz12} and references therein.

\section{Evaluation of correlation functions}\label{sec1}

The properties of plasmas are expressed in terms of correlation functions in thermodynamic equilibrium. Examples are thermodynamic properties (\ref{fik}) and transport properties (\ref{Kubo}).
There are several methods to calculate these correlation functions.
Exact solutions are known only for ideal quantum gases where there is no interaction potential $V$. The equations of state are known, e.g., the pressure $P$ is expressed by Fermi integrals. 
At fixed temperature, the equation of state for ideal classical gases $P=n k_BT$ is approximated by considering the limiting case of low density.
For electrical conductivity, $\sigma=\infty$ is obtained because of conservation of total momentum. The resistivity follows as $\rho=1/\sigma=0$ for charged ideal Fermi gases.

Correlations appear for the plasma Hamiltonian with complete interaction $V$. No closed-form solutions are known, and we must perform approximations to solve this many-body problem.
Here we discuss three possibilities:
\begin{enumerate}
\item 
Perturbation expansion with respect to $V$. We obtain analytic expressions for arbitrary orders of $V$ in terms of noninteracting equilibrium correlation functions, 
which can be easily evaluated using Wick's theorem. However, we have no proof of the convergence of this series expansion and no error estimate.
In order to make this analytical approach more efficient, the method of thermodynamic Green's functions and Feynman diagram technique 
were elaborated \cite{FW,GRBuch,KKER}.
Convergence is improved by performing partial summations corresponding to special concepts such as the introduction of the quasiparticle picture (self-energy $\Sigma$),
 screening of the potential (polarization function $\Pi$), or formation of bound states (Bethe-Salpeter equation). 
This leads to useful results for the properties of the plasma in a wide range of $T$ and $n$. 
However, as characteristic for perturbative approaches, exact results can be found only in some limiting cases.

\item

This drawback is eliminated by numerical simulations of the correlation functions that apply to arbitrary interaction strength. 
In Born-Oppenheimer approximation, density functional theory (DFT) for the electron system with given ion configuration and molecular dynamics (MD) for the ion system are applied to evaluate the correlation functions.
Single-electron states are calculated by solving the Kohn-Sham equations. The total energy is obtained from the kinetic energy of a non-interacting reference system, the classical electron-electron interaction, and an exchange-correlation energy that includes, to a certain approximation, all unknown contributions. 

The DFT-MD approach has been successfully applied to calculate the thermodynamic properties of complex materials in a wide range of $T$ and $n$, which will not be reported here, see, e.g., \cite{Lorenzen10,Wang13,Bonitz20,Tirelli22} and the references given there. 
For electrical conductivity (\ref{Kubo}), the Kubo-Greenwood formula \cite{Kubo66,Greenwood}
\begin{equation}
\label{eq:KG}
\operatorname{Re} \left[\sigma (\omega)\right] = \frac{2 \pi e^2}{3 m_e^2 \omega \Omega} \sum_k w_k \sum_{j=1}^N \sum_{i=1}^N  \sum_{\alpha=1}^3 
\big[ f(\epsilon_{j,k})- f(\epsilon_{i,k}) \big]  |\langle{\Psi_{j,k}| \hat{p}_\alpha | \Psi_{i, k}}\rangle|^2 \delta(\epsilon_{i,k} -\epsilon_{j,k} - \hbar\omega )
\end{equation}
was used to calculate the frequency-dependent dynamic electrical conductivity $\sigma(\omega)$ in the long-wavelength limit~\cite{Desjarlais02,Mazevet05,Holst11,French2017,Gajdos2006,RSRB21}. Kohn-Sham wave functions $\Psi_{i,k}$ from density functional theory calculations are used to calculate the transition matrix elements of the momentum operator $\hat{p}_\alpha$. The Fermi-Dirac distribution $f(\epsilon)$ accounts for the average occupation at energy $\epsilon$, and the summation over momentum space $k$ contains the $k$-point weights $w_k$.

Due to the finite size of the simulation box, the delta function in equation~\eqref{eq:KG} must be approximated by a finite-width Gaussian, which also prevents the direct calculation of the dc conductivity at $\omega = 0$. Therefore, the dynamic conductivity is extrapolated to the limit $\omega \rightarrow 0$ by a Drude fit,
\begin{equation}
\label{eq:Drude}
\operatorname{Re} \left[\sigma (\omega)\right] = \frac{ne^2 \nu}{\nu^2 + \omega^2},
\end{equation}
where $\nu$ is the collision frequency. Thus, the calculated direct current conductivity depends on choosing the appropriate width for the Gaussian and finding a suitable range for the Drude-fitting to $\sigma(\omega)$ calculated from equation~(\ref{eq:KG}). The last point can be improved by using a frequency-dependent collision frequency \cite{Heidi}.

One of the main shortcomings of the DFT-MD approach is that the many-particle interaction is replaced by a mean-field potential.  When using product wave functions for the many-electron system, correlations are excluded. The exchange-correlation energy density functional reflects the Coulomb interaction to some approximation, e.g., as it exists in the homogeneous electron gas,
but becomes problematic in the low-density limit where correlations are important.

\item
In principle, an accurate evaluation of equilibrium correlation functions is possible using path-integral Monte Carlo (PIMC) simulations, see~\cite{Dornheim2018,PIMC1,PIMC2} and references therein.
The shortcomings of this approach at present are the relatively small number of particles (a few dozen), the sign problem for fermions, and the computational challenges in accurately computing path integrals.
Instead of using an exchange-correlation energy density functional, $e-e$ collisions are treated accurately. However, at present accurate calculations have only been performed for the uniform electron gas model in which the charge-compensating ion subsystem is replaced by a homogeneously charged jellium. The results presented in \cite{TD} are shown below in sec. \ref{sec:4}. 
High-precision calculations for the two-component Hydrogen plasma would be of interest for both thermodynamics and transport properties.
\end{enumerate}

\section{Green's functions and Feynman diagrams}\label{sec2}

In quantum statistics, the method of thermodynamic Green's functions has been worked out to evaluate correlation functions in thermodynamic equilibrium.
For the ideal quantum gas, in which there is no interaction, all equilibrium correlation functions can be calculated using Wick's theorem. 
For plasmas, we can perform a power series expansion with respect to the interaction strength according to the Dyson series.
The terms of this perturbation expansion are represented by Feynman diagrams.

The problem of the perturbation expansion is that the convergence property remains open, and we cannot anticipate that for the correlation functions a power series expansion with respect to the interaction strength is possible.
A predetermined wrong analytical behavior near the singular case of ideal gases leads to divergencies which are avoided performing partial summations that can modify the analytic behavior.
The most important partial summations are the quasiparticle concept associated with the introduction of the self-energy, the screening associated with the introduction of the polarization function, and the introduction of bound states 
performing partial summation of ladder diagrams. For instance, the Bethe-Salpeter equation for the two-particle Green function in ladder approximation corresponds to the solution of the two-body problem.

From classical statistics, the Mayer cluster expansion is well known for short-range potentials  is well known for the partition function, and the virial expansion in powers in $n$ is obtained.
Because of the long-range nature of the Coulomb potential, this expansion in powers in $n$ is not possible for plasmas, the virial coefficients are divergent.
Screening, i.e. partial summation of the so-called ring diagrams in quantum statistics, solves this convergence problem, and the expansion in powers of $n^{1/2}$ is possible.
When considering the spectral function, the contribution of the free particles is replaced by the contribution of the quasiparticles, with the energies containing the Debye shift.
To obtain the thermodynamic potentials $F$ or $P\Omega$ from the equation of state (\ref{EOS}) we must perform integration over $\mu$ or $n$, respectively,
and logarithmic terms may appear. In particular, for the free energy of the Hydrogen plasma, the virial expansion reads
\begin{eqnarray}
\label{Fvir}
&&F(T,\Omega,N)=\Omega k_BT \left\{n \ln n + [3/2\ln(2 \pi \hbar^2/(m k_BT))-1] n
\right. 
\nonumber \\ &&\left.-F_0(T)n^{3/2}-F_1(T)n^2 \ln n-F_2(T) n^2
%\right. \nonumber \\ &&\left. 
-F_3(T)n^{5/2} \ln n-F_4(T) n^{5/2}+{\cal O}(n^3\ln n)\right\}.
\end{eqnarray}
see  \cite{KKER,TD} where expressions for the lowest virial coefficients $F_i$ are also given. Details on the calculation of the EoS for Coulomb systems can be found in Ref. \cite{KKER} and will not be repeated here. 
The virial expansion for the uniform electron gas is discussed below in Sec. \ref{sec:4}.

Perturbation expansion and partial summations also apply to the evaluation of the correlation function (\ref{Kubo}) which is related to the electrical conductivity.
In the lowest order of perturbation theory, where interactions are neglected, the total momentum of the electrons is conserved. 
As a consequence, the expression (\ref{Kubo}) becomes divergent, the ideal plasma shows no finite value for the conductivity.
Partial summations, in particular the self-energy and vertex corrections, lead to finite values for the conductivity, see \cite{Rcond2018}.
Analytical evaluation of the Kubo formula remains  difficult and cumbersome.

In contrast, it is possible to perform a virial expansion for the inverse conductivity $R=1/\sigma$, expressed as a correlation function of the stochastic forces \cite{Rcond2018}. 
A generalized linear response theory was worked out that takes into account  correlation functions of higher moments of the occupation number distribution \cite{Roep88}.
In this way the relation to the kinetic theory was shown \cite{Heidi}. 
These correlation functions are also treated by the methods of Green functions, Feynman diagram techniques and partial summations, so that virial expansions can be carried out.

The dc conductivity $\sigma(n,T)$ is usually associated with a dimensionless function $\sigma^*(n,T)$ according to 
\begin{eqnarray}
 \sigma(n,T) &=& \frac{(k_BT)^{3/2} (4\pi\epsilon_0)^2}{m_e^{1/2} e^2}\;\sigma^*(n,T). 
 %\nonumber\\&=&\frac{32405.4}{\Omega {\rm m}}\left(\frac{k_BT}{\rm eV}\right)^{3/2} \sigma^*(n,T) \,.
 \label{eq:1}
\end{eqnarray} 
We consider both $\sigma$ and $\sigma^*$ as a function of density $n$ at {\it fixed} temperature $T$. In the limiting case of low density, the following virial expansion for the inverse conductivity $\rho^*(n,T)=1/\sigma^*(n,T)$ was obtained from kinetic theory and generalized linear response theory~\cite{Roep88,RR89,Redmer97}: 
\begin{equation}
 \rho^*(n,T) = \rho_1(T) \ln\frac{1}{n} + \rho_2(T) + \rho_3(T)\,n^{1/2}\,\ln\frac{1}{n} + {\cal O}(n^{1/2}),
 \label{eq:5}
\end{equation}
which begins with a logarithmic term. Values for the virial coefficients $\rho_i(T)$ are given below in Sec. \ref{sec:5}.

\section{Virial plots}\label{sec3}

Equilibrium properties, such as the correlation functions considered here, depend on a limited number of state variables.
For the Hydrogen plasma, this are the temperature $T$ and the electron number density $n$
(for charge neutral plasmas, the ion (proton) number density is also $n$).
For the uniform electron gas, we have the same variables. Instead of the ion subsystem a homogeneously charged background (jellium model) 
is considered
to establish charge neutrality. 
In the case of a many-component plasma, the independent partial densities $n_i$ (not connected by chemical reactions and charge neutrality) of the components are the state variables in addition to $T$. 
We focus here on the two simple cases where the state variables are $T,n$, and we study the correlation energy $\bar V(T,n)$ of the uniform electron gas
and the electrical conductivity $\sigma(T,n)$ of the Hydrogen plasma, in particular the resistivity $R(T,n)=1/\sigma(T,n)$.

It is convenient to introduce dimensionless variables instead of $T,n$.
We use atomic units with the Hartree energy
\begin{equation}
 E_{\rm Ha}=\left(\frac{e^2}{4 \pi \epsilon_0}\right)^2 \frac{m}{\hbar^2}=27,21137\,{\rm eV} =2\, {\rm Ry}
\end{equation}
and the Bohr radius
\begin{equation}
 a_B=\frac{4 \pi \epsilon_0}{e^2} \frac{\hbar^2}{m}=5.2918 \times 10^{-11}\,{\rm m}.
\end{equation}
The density in atomic units is usually represented by the radius of a sphere containing an electron,
\begin{equation}
 r_s= \left(\frac{3}{4 \pi n}\right)^{1/3}\frac{1}{a_B}.
\end{equation}
The temperature is related to the energy $k_BT$, so that  1~eV corresponds to 11604.6~K.
We denote $T_{\rm eV}$ as $k_BT$ measured in units of eV, $T_{\rm Ha}$ in units of $E_{\rm Ha}$, and $T_{\rm Ry}$ in units of Ry so that
\begin{equation}
 T_{\rm Ha}=\frac{k_BT}{E_{\rm Ha}}= 2 T_{\rm Ry}=27,21137\, T_{\rm eV}.
\end{equation}
Another well-known choice of dimensionless parameters is
\begin{equation}
 \Gamma = \frac{e^2}{4\pi\epsilon_0 k_BT} \left(\frac{4\pi}{3} n \right)^{1/3},\qquad  \Theta=\frac{2mk_BT}{\hbar^2}(3 \pi^2 n)^{-2/3}.
\end{equation}
The plasma parameter $\Gamma$ characterises the ratio of potential to kinetic energy in the non-degenerate case, 
and the electron degeneracy parameter $\Theta$ characterises the range in which the electrons are degenerate.
Different sets of dimensionless parameters are related. 
Thus, PIMC calculations for specific parameter values of $r_s, \Theta$ are discussed in the following section,
the corresponding plasma parameters $n,T$ are determined as follows,
\begin{equation}
 n= \frac{3}{4 \pi}\frac{1}{(r_sa_B)^3},\qquad k_BT=E_{\rm Ha}\frac{1}{2}\left(\frac{9 \pi}{4}\right)^{2/3} \frac{\Theta}{r_s^2}
\end{equation}
with $E_{\rm Ha}/k_B= 315777.1$ K.

The dc conductivity $\sigma(n,T)$ is also associated with a dimensionless function 
$\sigma^*(n,T)$ according to 
\begin{equation}
\sigma(n,T) = \frac{(k_BT)^{3/2} (4 \pi \epsilon_0 )^2}{m_e^{1/2} e^2}\;\sigma^* = 
 0.0258883\,\, T^{3/2}\;\sigma^* (\Omega {\rm m\,\,K}^{3/2})^{-1} = 
 32405.4\,\, T_{\rm eV}^{3/2}\;\sigma^* (\Omega {\rm m})^{-1}\,.
\end{equation}
As with thermodynamic relations, the dimensionless conductivity $\sigma^*$ can be expressed as a function of dimensionless 
variables $r_s, T_{\rm Ha}$ or $\Gamma,\Theta$.  These functions are now to be specified. 
Exact results are currently known only for limiting cases, in particular virial expansions.

The analysis of a virial expansion is sometimes not  easy because trivial terms dominate in limiting cases so that interesting terms remain hidden.
In the example of the thermodynamic EoS considered in Sec. \ref{sec:4}, one dominant term is the Debye shift, which covers the contribution of higher virial coefficients.
We introduce reduced virial expansions where these exactly known contributions are suppressed, 
and quantities are introduced that anticipate a linear relationship in special cases.
The virial plot is the representation of this asymptotic linear relationship and allows us to extrapolate virial coefficients from simulations. 
We demonstrate this procedure for two cases, the mean potential energy of the uniform electron gas in Sec. \ref{sec:4} and the electrical conductivity 
of the Hydrogen plasma in Sec. \ref{sec:5}.

If we express $\sigma^*(n,T)$ in terms of dimensionless parameters $\Gamma,\Theta$ and
use the Born parameter $\Gamma/\Theta$, which is of interest in the range $k_BT \gg 1$ Ry,
from Eq. (\ref{eq:5}) we obtain a modified virial expansion where the argument of the logarithm is dimensionless,
\begin{eqnarray}
&&\frac{1}{\sigma^*(\Gamma,\Theta)} = \rho^*(\Gamma, \Theta)= \tilde \rho_1(\Gamma^2 \Theta)\ln\left(\frac{\Theta}{\Gamma}\right) + 
 \tilde \rho_2(\Gamma^2 \Theta)+\dots\,,\nonumber \\
 &&\Gamma^2 \Theta=\frac{2^{7/3}}{3^{4/3}\pi^{3/3}} \frac{1}{T_{\rm Ha}},\qquad
 \frac{\Theta}{\Gamma}=\frac{2^{1/3}}{3^{1/3}\pi^{5/3}} \frac{T^2_{\rm Ha}}{n a_B^3}
\label{eq:5a}
\end{eqnarray}
We define the reduced effective virial coefficient $\tilde\rho^{\rm eff}_2(T)$ according to
\begin{equation}
 \tilde\rho^{\rm eff}_2(n,T) = \frac{32405.4}{\sigma (n,T) [\Omega {\rm m}]} T_{\rm eV}^{3/2} 
 -  \tilde \rho_1(T) \,\, \ln\left(\frac{\Theta}{\Gamma}\right),
 \label{rho2eff1}
\end{equation}
with $\lim_{n \to 0}\tilde\rho^{\rm eff}_2(n,T)=\tilde\rho_2(T)$, see also Eq.~(\ref{rho2eff}) below. The plot of $ \rho^*/\ln(\Theta/\Gamma)$ as a function of $x=1/\ln(\Theta/\Gamma)$ at given $T$ is called a virial plot. It directly allows the determination the virial coefficients $\rho_1(T), \rho_2(T)$, as it is shown in Sec. \ref{sec:5}.
%%%%%%%%%%%%%%%%%%%%%%%

As will be demonstrated in this work, virial plots are very sensitive to diverse approaches, including the results of numerical simulations,
in the low density domain. Since trivial dominant terms, which are known exactly, are suppressed, they have no effects due to possible approximations,
and the extrapolation of numerical simulations into the low-density domain becomes immediately possible.

\section{Virial expansion of the EoS of the UEG, comparison with PIMC simulations}\label{sec:4}

%\subsection{Virial Expansion and coefficients from PIMC for the uniform electron gas}

The problem of the second virial coefficient  for the mean correlation energy $\bar V$ was considered in a recent work \cite{TD}.  
There was a controversy about the high-temperature limit of the second virial coefficient, i.e. the term $\propto 1/\sqrt{T}$ \cite{KKR15}.
This controversy disappears in charge-neutral two-component plasmas, but not in the uniform electron gas (UEG), where interacting electrons 
are moving in front of a positively charged jellium-like background to neutralize the Coulomb field at large distances. Accurate PIMC simulations have been available
at low densities and high temperatures \cite{TD}, so that it was possible to confirm the correct limiting behavior. 
In this section, we not only show the virial plot method to confirm the correct limiting law, but consider the full second virial coefficient and discuss deviations 
from this expansion.

The virial expansion of the free energy $F(T,\Omega,N) $ of the UEG is obtained from the general formula for a multi-component plasma given in \cite{KKER,TD}.
The mean potential energy $V$ is  determined by
\begin{equation}
V(T,\Omega,N)=e^2 \frac{\partial}{\partial (e^2)} F(T,\Omega,N)
\end{equation}
(for the relation to the internal energy see \cite{Kraeft02}). 

From the virial expansion of  $F(T,\Omega,N) $, we get the following virial expansion of $V$
\begin{eqnarray}
&&\frac{V}{Nk_BT}=-\frac{\kappa^3}{8 \pi n}- \pi n\lambda^3 \tau^3 \ln(\kappa\lambda)\nonumber \\
&&-\pi n \lambda^3\left[\frac{\tau}{2}-\frac{\sqrt{\pi}}{2} (1+\ln(2))\tau^2+\left(\frac{C}{2}+\ln(3)-\frac{1}{3}+\frac{\pi^2}{24}\right)\tau^3\right. \nonumber \\
&&\left. +\sqrt{\pi}\sum_{m=4}^\infty\frac{(-1)^mm}{2^m \Gamma(m/2+1)}\left[2 \zeta(m-2)-(1-4/2^m)\zeta(m-1)\right]\tau^m \right] \nonumber \\
&&- \pi n\lambda^4 \tau^4 \kappa \ln(\kappa\lambda)+\frac{V_4(T)}{Nk_BT}n^{3/2} +{\cal O}(n^2 \ln(n))
\label{virV}
\end{eqnarray}
with the variables
\begin{equation}
\kappa^2 = \frac{ne^2}{\epsilon_0k_BT},\qquad \lambda^2=\frac{\hbar^2}{m k_BT}, \qquad \tau = \frac{e^2 \sqrt{m}}{4 \pi \epsilon_0 \sqrt{k_BT}\hbar}.
\end{equation}
$\zeta(x)$ denotes the Riemann zeta function, and $C=0.57721\dots$ is Euler's constant. We express this expansion in terms of $T, n$ and introduce atomic units $\hbar=m=e^2/4 \pi \epsilon _0=1$ 
so that $k_BT$ is measured in Hartree (Ha) and $n$ in electrons per $a_B^3$.

The virial expansion of the specific mean potential energy $v=V/N$ is as follows
\begin{equation}
v(T,n)=v_0(T) n^{1/2}+v_1(T) n \ln\left(\kappa^2 \lambda^2\right)+v_2(T) n+v_3(T)n^{3/2} \ln\left(\kappa^2 \lambda^2\right)+v_4(T) n^{3/2}+{\cal O}(n^2 \ln(n)).
\end{equation}
If atomic units are used, this results in ($\kappa^2 \lambda^2=4 \pi n/T^2$)
\begin{eqnarray}
v_0(T)&=&-\frac{\sqrt{\pi}}{T^{1/2}}, \qquad v_1(T)=-\frac{\pi}{2 T^2},\nonumber\\
v_2(T)&=&-\frac{\pi}{T}\left[\frac{1}{2}-\frac{\sqrt{\pi}}{2}(1+\ln(2))\frac{1}{T^{1/2}}+\left(\frac{C}{2}+\ln(3)-\frac{1}{3}+\frac{\pi^2}{24} \right) \frac{1}{T}\right.\nonumber \\
&&\left. - \sqrt{\pi}\sum_{m=4}^\infty\frac{m}{2^m \Gamma(m/2+1)}\left(\frac{-1}{T^{1/2}}\right)^{m-1} [2 \zeta(m-2)-(1-4/2^m)\zeta(m-1)]\right],\nonumber \\
v_3(T)&=&-\frac{3 \pi^{3/2}}{2 T^{7/2}}.
\label{v0123}
\end{eqnarray}

In ref. \cite{TD}, a virial plot was presented to study the behavior of the second virial coefficient. 
We consider the lowest orders of the virial expansion,
\begin{equation}
v^{(1)}(T,n)=-\frac{\sqrt{\pi}}{T^{1/2}} n^{1/2}-\frac{\pi}{2 T^2} n \ln\left(\frac{4 \pi n}{T^2}\right),
\end{equation}
as exactly known and subtract them from the data obtained from the PIMC simulations, $v^{\rm PIMC}=V^{\rm PIMC}/N$.
These exactly known terms may become very large, hiding the higher virial coefficients. (Note that the logarithmic term contains a factor to become dimensionless.
This factor can be moved to the next virial coefficient.) 

In \cite{TD} we introduced the reduced potential energy ($\tau =T^{-1/2}$, atomic units)
\begin{eqnarray}
&&v_2^{\rm red}(T,n)=\left[v^{\rm PIMC}-v^{(1)}(T,n)\right] \frac{-T}{\pi n}=\frac{-T}{\pi}v_2(T)+{\cal O}(n^{1/2}\ln(n))\nonumber\\
&&=\frac{1}{2}-\frac{\sqrt{\pi}}{2} (1+\ln(2))\tau+\left(\frac{C}{2}+\ln(3)-\frac{1}{3}
+\frac{\pi^2}{24}\right)\tau^2+{\cal O}(\tau^3)+{\cal O}(n^{1/2}\ln(n)).
%\nonumber\\
%&&=\frac{-T}{\pi}v_2(T)+{\cal O}(n^{1/2}\ln(n)).
%+{\cal O}(n^{1/2}\ln(n)).
\label{v2vir}
\end{eqnarray}
%if higher orders of density are neglected.

\begin{table}[h]
\begin{center}
\caption{PIMC calculations for the uniform electron gas: $v^{\rm PIMC}$ and  $v_2^{\rm red}$, eq. (\ref{v2vir}), for special parameter values $r_s, \Theta$ and the corresponding values of $T, \tau, n$.\label{tab:1}}
 \begin{tabular}{|c|c|c|c|c|c|c|c|}
%\hspace{0.5cm}
%\hline
\toprule
$r_s$& $\Theta$ & $v^{\rm PIMC}$ [Ha]  &  $T_{\rm Ha}$ & $\tau$& $v_2^{\rm red}$ & $T$ [K]  & $n$ [cm$^{-3}$]  \\ 
\hline
0.5&128    &-0.0826214    &942.891    & 0.0325664	& 0.453524 &2.97742e8   & 1.28882e25 \\
    &64       &-0.1180456      &471.446    &0.0460558	& 0.420822	&1.48871e8    &1.28882e25 \\
    &32       &-0.169272      &235.723    &0.0651327	& 0.398701	&7.44354e7    & 1.28882e25 \\
    &16     	&-0.2423993       &117.861    &0.0921116	& 0.356465	&3.72177e7   & 1.28882e25 \\
    &8      	&-0.3447641         &58.9307    &0.130265	& 0.294433	&1.86089e7   & 1.28882e25 \\
\hline
2	& 128	& -0.0402248	&58.9307		&0.130265		&0.290766	&1.8609e7		&2.01378e23 \\
	& 64		& -0.0568062	&29.4653		&0.184223		&0.257047	&9.30448e6	&2.01378e23 \\
	& 32		& -0.0797147 	&14.7327		&0.260531		&0.207038	&4.65224e6	&2.01378e23\\
	& 16		& -0.1101257		&7.36634		&0.368446		&0.126496	&2.32612e6	&2.01378e23\\
	&  8		& -0.1486611		&3.68317		&0.521062		&0.0596564	&1.16306e6	&2.01378e23\\
	\hline
20& 128		& -0.0119299		&0.589307	&1.30265			&1.50247			&186090.		&2.01378e20\\
	& 64		& -0.0160051	&0.294653	&1.84223			&3.48031			&93044.8		&2.01378e20\\
	& 32		& -0.0207112		&0.147327	&2.60531			&6.67878			&46522.4		&2.01378e20\\
	& 16		& -0.0256337	&0.0736634	&3.68446 	&10.2475			&23261.2		&2.01378e20\\
	&  8		& -0.0302098	&0.0368317	&5.21062		&9.50255			&11630.6		&2.01378e20\\
\hline
 \end{tabular}
\end{center}
\end{table}

\begin{figure}[t]
\centerline{\includegraphics[width=0.7 \textwidth]{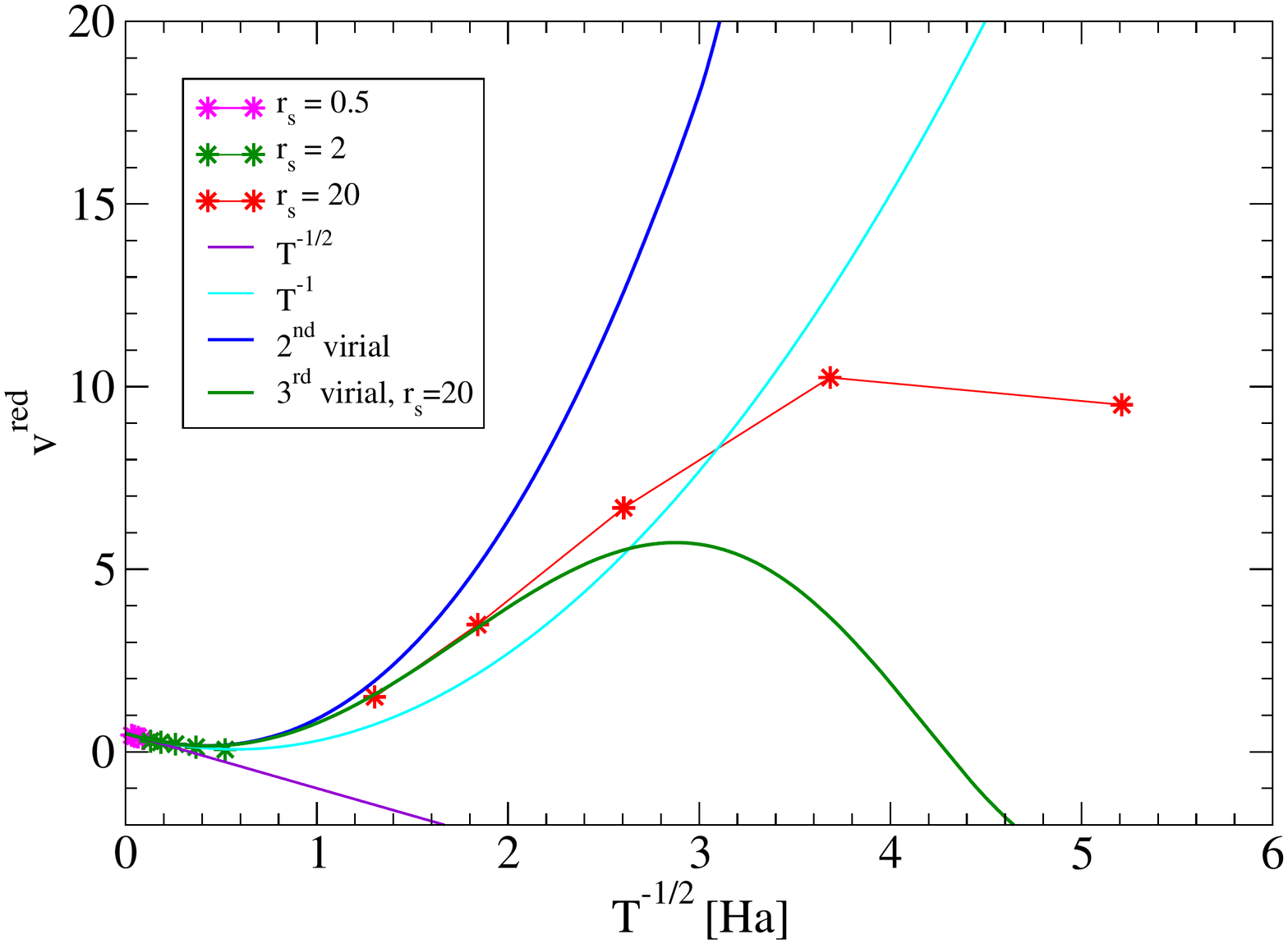}}
\caption{Reduced potential energy $v_2^{\rm red}(T,n)$, Eq. (\ref{v2vir}), as function of $\tau = 1/\sqrt{T}$ for different densities, $r_s=0,5; 2; 20$. 
For comparison, the reduced second virial coefficient $v_2^{\rm red}(T)=-(T/\pi) v_2(T)$ [2$^{\rm nd}$ virial, according Eq.  (\ref{v0123})] as well as the lowest orders 
in $1/T$ are shown. In addition, the curve 3$^{\rm rd}$ virial given by Eq. (\ref{v2vi3}) is also shown.  (Atomic units are used.)\label{fig:1}}
\end{figure}
\begin{figure}[t]
\centerline{\includegraphics[width=0.7 \textwidth]{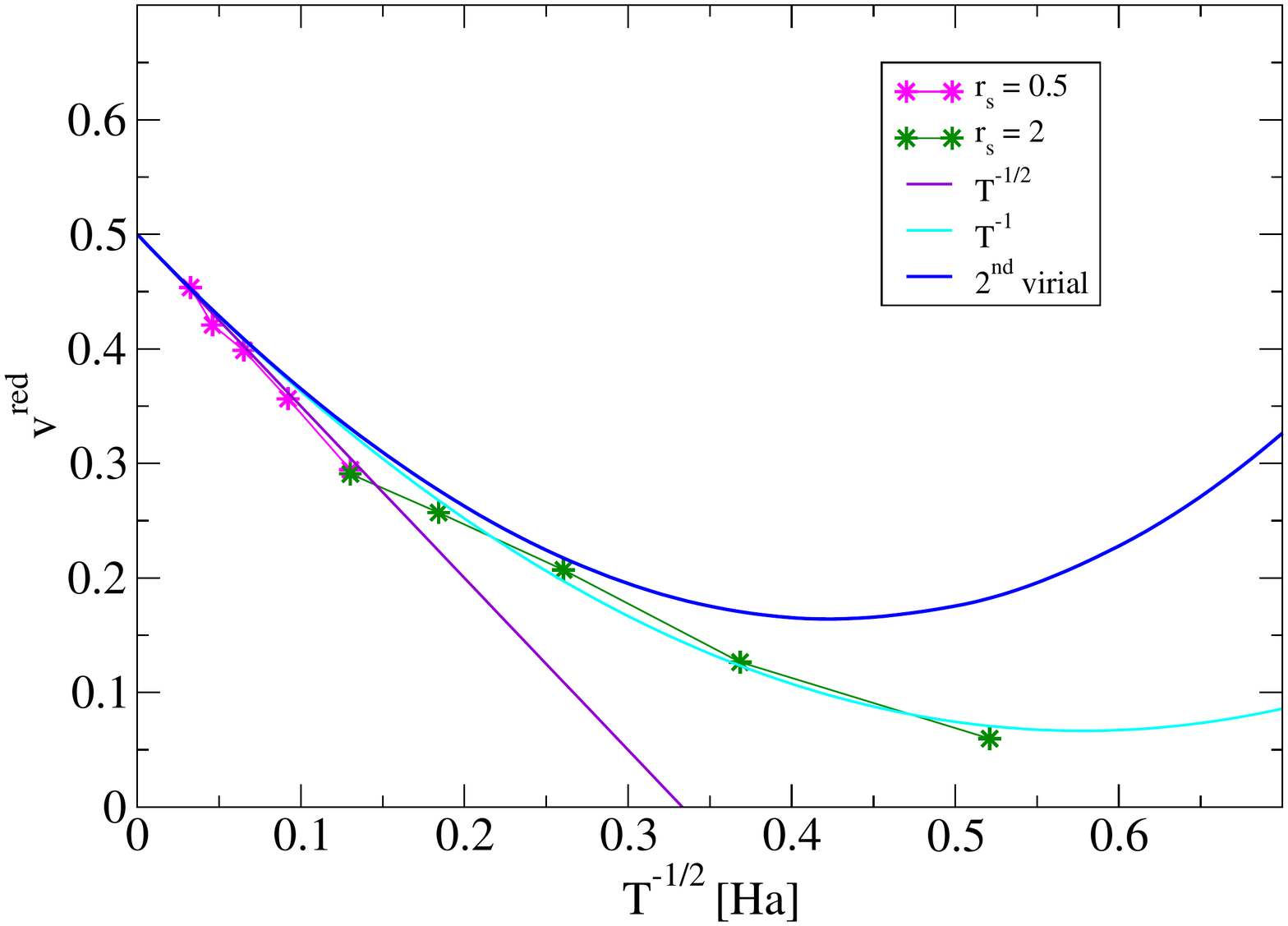}}
\caption{Detail of Fig. \ref{fig:1}.\label{fig:2}}
\end{figure}

In Tab. \ref{tab:1}, the parameter values of the uniform electron gas are given for which PIMC calculations were presented in Ref. \cite{TD}, 
together with the values for $v_2^{\rm red}$ (\ref{v2vir}). The results for $v_2^{\rm red}$ are also shown in Figs. \ref{fig:1}, \ref{fig:2}.

In Fig. \ref{fig:1} all calculated PIMC data of \cite{TD} are considered and the corresponding value of $v_2^{\rm red}$ is shown as function of $\tau$, see Tab. \ref{tab:1}.
In addition, three expressions for (\ref{v2vir}) are shown: up to order $\tau$, i.e., $1/2-\sqrt{\pi} (1+\ln(2))\tau/2$, up to order $\tau^2$, and the full $\tau$ dependence.
This Fig. \ref{fig:1} shows in which interval of $\tau$ the linear or quadratic approximation is applicable. 
The PIMC data are very different. The lowest density, $r_s=20$, should be most relevant to the low-density limit, where higher virial coefficients are less important.
However, the inverse temperature $\tau=T_{\rm Ha}^{-1/2}$ is too large to reach the limit $\tau \to 0$. 
Close to this limit are PIMC simulation data for $r_s=0.5$. The relatively large density is compensated by the very high temperature, see Tab. \ref{tab:1}.

A part of Fig. \ref{fig:1} is shown enlarged in Fig. \ref{fig:2}. It was a main result of Ref.  \cite{TD} to show that the PIMC simulation data confirm the limit $v_2^{\rm red}(\tau=0)=1/2$.
Linear fit to the data for  $r_s=0.5$ is possible, and extrapolation to $v_2^{\rm red}(\tau=0)$ gives 1/2. 
At the same time, one gets an idea of the accuracy of the simulation, which shows up as scatter around the analytical behavior.
The PIMC data for $r_s=2$ are not described by the linear approximation but almost well by the quadratic approximation. 
Finally, we have to make a comparison with the full second virial coefficient and will find that good agreement is obtained in all three density cases, given by the parameter $r_s$, 
only for the lowest values of $\tau$ (an exception is the lowest $\tau$ parameter calculation for $r_s=2$, which needs to be checked).
As $\tau$ increases, the PIMC data are systematically below the second virial curve. 
We assume that the PIMC simulations are very accurate, so this deviation indicates the contribution of higher virial coefficients. 

Deviations from the second virial coefficient $-(T/\pi) v_2(T)$ indicate the contribution of higher orders to the virial expansion. 
We expect a significant next order contribution from the low-density calculations, i.e., $r_s=20$. We consider the expression
\begin{equation}
v_{2+3}^{\rm red}(T,n) =-\frac{T}{\pi}\left[v_2(T)+v_3(T)n^{1/2}\ln\left(\frac{4 \pi n}{T^2}\right)\right],
\label{v2vi3}
\end{equation}
which accounts for the contribution of the third virial coefficient. For $r_s=20$, the data are well reproduced for the lowest values of $\tau$, see also Fig. \ref{fig:1}. 
Deviations for larger $\tau$ indicate the contributions of higher virial coefficients.

The deviation 
\begin{equation}
\Delta v^{\rm red}_2(T,n)=\left[v^{\rm PIMC}-v^{(1)}(T,n)-v_2(T)n\right]\,\frac{T}{\pi n}
\end{equation}
is shown in Tab. \ref{tab:2}, together with the deviation
\begin{equation}
\Delta v^{\rm red}_3(T,n)=\left[v^{\rm PIMC}-v^{(1)}(T,n)-v_2(T)n-v_3(T)n^{3/2} \ln\left(\frac{4 \pi n}{T^2}\right)\right]\,\frac{T}{\pi n}.
\end{equation}
As mentioned  before,
the inclusion of the third virial coefficient $v_3(T)$ improves the agreement of the PIMC simulations with the virial expansion, as also shown in Fig. \ref{fig:1}. 
The remaining difference $\Delta v^{\rm red}_3(T,n)$ is related to the fourth-order and higher-order virial coefficient,
\begin{equation}
v_4^{\rm eff}(T,n)=\Delta v^{\rm red}_3(T,n)\,\frac{\pi}{T n^{1/2}}=v_4(T)+{\cal O}(n^{1/2} \ln(n)).
\label{eq:v4}
\end{equation}
The fourth virial coefficient results when higher-order virial coefficients are neglected, $\lim_{n \to 0}v_4^{\rm eff}(T,n)=v_4(T)$.
This should be possible in the low-density limit, where the contributions of higher orders of the density expansion become small. 
However, high-precision calculations are required to extract the higher-order coefficients, 
and the accuracy of the present calculations \cite{TD} is not sufficient to determine precisely the fourth- and higher-order virial coefficients.
We give here only a discussion of the present data.

\begin{figure}[t]
\centerline{\includegraphics[width=0.7 \textwidth]{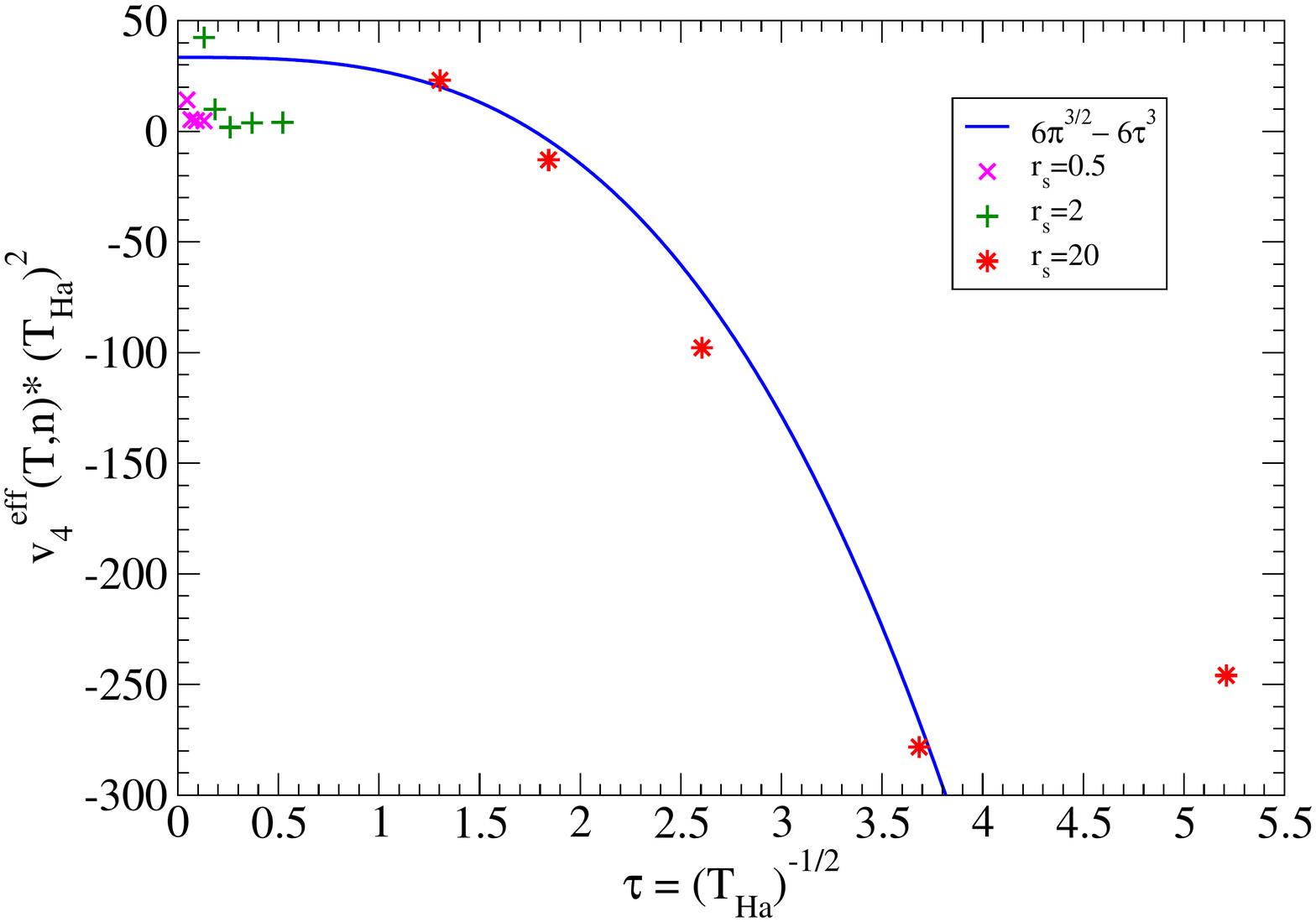}}
\caption{Effective reduced fourth virial coefficient $v_4^{\rm eff}(T,n) \times T^2$, Eq. (\ref{eq:v4}), 
plotted as function of $\tau = 1/\sqrt{T_{\rm Ha}}$ for different densities, $r_s=0,5; 2; 20$. 
For comparison, a curve 
%$5 - 5 \times \tau^3$ 
$3 \pi \sqrt{4 \pi} - 6  \tau^3$ 
is seen. (Atomic units used.) \label{fig:2a}}
\end{figure}

From the virial expansion of the free energy \cite{KKER},  the fourth virial coefficient  $v_4(T)$ contains contributions 
with temperature dependence $\propto T^{-2}=\tau^4$ and higher orders in $\tau$, as well as contributions $\propto T^{-7/2}$. 
The coefficient  of the $\tau^4$ term follows as $3 \pi \sqrt{4 \pi}$. 
We expect a high-temperature limit behavior $\propto T^{-2}$, and we
show in Fig. \ref{fig:2a} the quantity $v_4^{\rm eff}(T,n) \times T^2$.

We see that the lowest value of density, $r_s=20$, exhibits behavior at small $\tau$ values  that can be compared to a curve $3 \pi \sqrt{4 \pi} - 6 \times \tau^3$.  However, the exact determination of the fourth virial coefficient $v_4(T)$ is not possible from the available data.
At the higher densities corresponding to smaller $r_s$, the accuracy of the numerical PIMC simulations may not be sufficient to extract higher-order virial coefficients.
In the context of our analysis, in addition to the dependence on $T$, the dependence on $n$ would be of interest to perform the virial plot as a function of $n$. Further calculations 
for density parameter values in the range of $r_s=20$ would be required. Since we are investigating the differences between large numbers, high accuracy is necessary.

The study of the uniform electron gas is not only of interest for the discussion of the exchange-correlation term of the energy-density functional in DFT calculations, for which Dornheim, Groth, and Bonitz derived analytical formulas \cite{DGB2018,GDB2017}.
It is also a prerequisite to treat the more interesting case of a two-component plasma, e.g., the Hydrogen plasma.
The equation of state at low densities is of interest, for example, in helioseismology \cite{Daeppen1988}, where the fourth virial coefficient $v_4(T)$ is important \cite{DeWitt1998}.
In this context, the high-temperature limit of $v_2^{\rm red}(\tau=0)$ was discussed in \cite{KKR15,TD}. 
For a discussion of the fourth virial coefficient $v_4(T)$ of Hydrogen plasma, see also Alastuey and 
Ballenegger \cite{AB10,AB12}.

\begin{table}[h]
\begin{center}
\caption{PIMC calculations for the UEG: $v$ and $v^{\rm red}$. The calculation with the second virial coefficient, Eq. (32), is denoted by $v_{\rm vir}$ and $v^{\rm red}_{\rm vir}$.\label{tab:2}}
 \begin{tabular}{|c|c|c|c|c|c|c|c|c|}
%\hspace{0.5cm}
%\hline
\toprule
$r_s$& $\Theta$ & $v$ [Ha]  &  $T_{\rm Ha}$& $n \,a_B^{3}$  & $\tau$& $ v^{\rm red}_2$ & $\Delta v^{\rm red}_2$ &$\Delta v^{\rm red}_3$  \\ 
\hline
0.5&128  &-0.082621    &942.891    & 1.90986	& 0.0325664	& 0.453524	&-0.000818266  &-0.000819682 \\
    &64     &-0.118045       &471.446    & 1.90986	&0.0460558	&  0.420822		&0.0132298 &0.0132228 \\
    &32     &-0.169272      &235.723    & 1.90986	&0.0651327	& 0.398701	&0.00992642  &0.00989306\\
    &16     	&-0.242399       &117.861    & 1.90986	&0.0921116	& 0.356465	&0.0181564  &0.0180015\\
    &8      	&-0.344764         &58.9307    & 1.90986	&0.130265	& 0.294433	&0.0360992  &0.0354136\\
\hline
2	& 128	& -0.040224	&58.9307		& 0.0298416	&0.130265		&0.290766	&0.039767		&0.0396097 	\\
	& 64		& -0.056806	&29.4653		& 0.0298416	&0.184223		&0.257047	&0.0194226 		&0.0186676 	\\
	& 32		& -0.079714 	&14.7327		& 0.0298416	&0.260531		&0.207038	&0.0103138 	&0.00680714\\
	& 16		& -0.110125	&7.36634		& 0.0298416	&0.368446		&0.126496	&0.043972 			&0.0284584	\\
	&  8		& -0.148661	&3.68317		& 0.0298416	&0.521062		&0.0596564	&0.12329 	&0.0599871	\\
	\hline
20	& 128	& -0.011929	&0.589307	& 0.0000298416	&1.30265		&1.50247			&0.440148 	&0.0680086 	\\
	& 64		& -0.016005	&0.294653	& 0.0000298416	&1.84223		&3.48031			&1.60138		& -0.0765312 	\\
	& 32		& -0.020711	&0.147327	& 0.0000298416	&2.60531		&6.67878			&5.91999		&-1.155		\\
	& 16		& -0.025633	&0.0736634	& 0.0000298416	&3.68446 		&10.2475			&19.7824		&-6.56867		 \\
	&  8		& -0.030209	&0.0368317	& 0.0000298416	&5.21062		&9.50255			&60.1201 		&-11.6087 	\\
\hline
 \end{tabular}
\end{center}
\end{table}

\section{Virial expansion of the inverse conductivity of H plasmas, comparison to DFT-MD simulations}\label{sec:5}

\begin{figure}[t]
\centerline{\includegraphics[width=0.7 \textwidth]{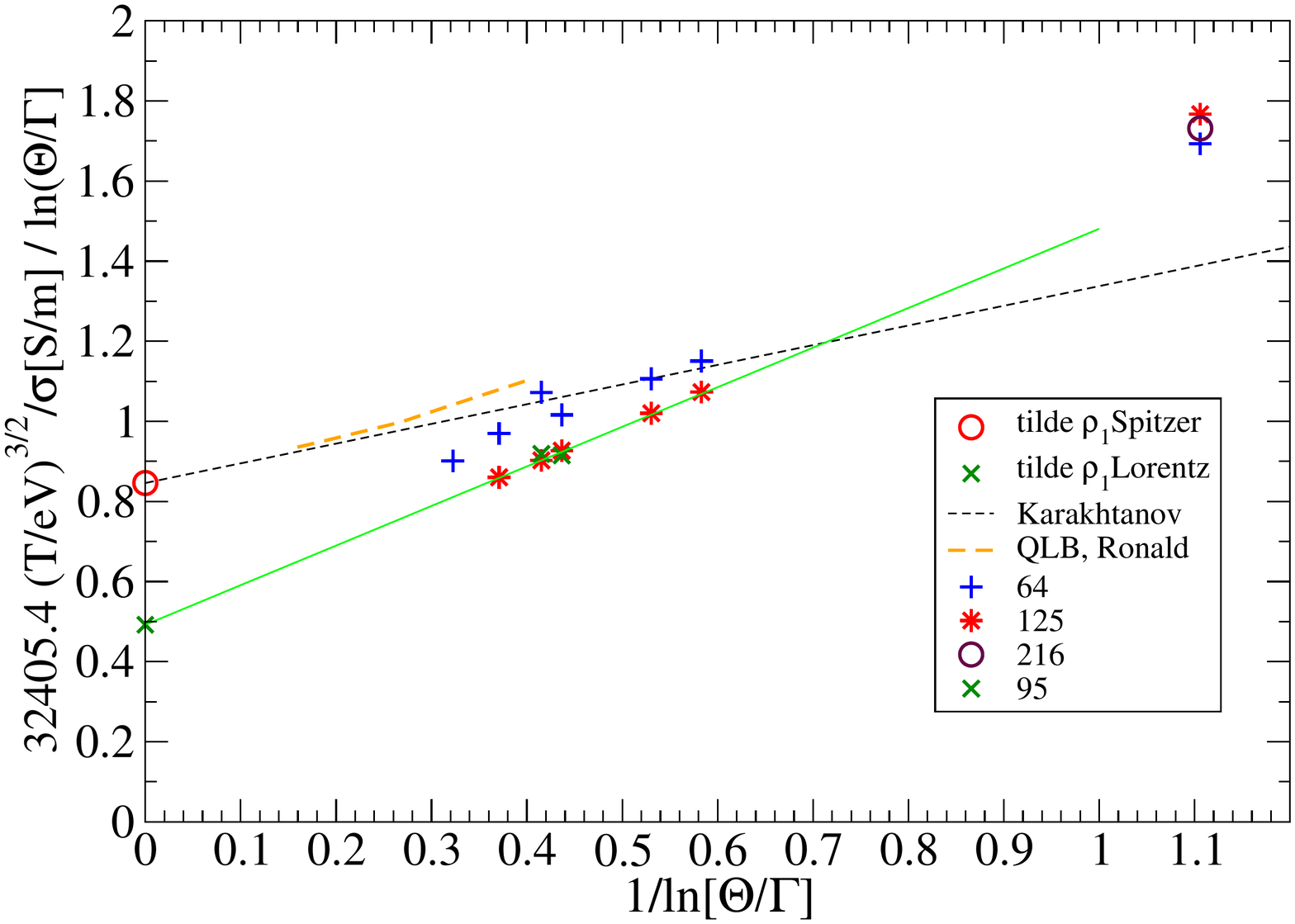}}
\caption{Reduced resistivity $\tilde\rho(x,T)$ (\ref{virrho}) for hydrogen plasma as a function of $x=1/\ln (\Theta/\Gamma)$: 
%QMD simulations of Lambert {\it et al.} \cite{Lambert11} (the orange line points to the value for $n=80$ g/cm$^3$, $k_BT=300$ eV), 
DFT-MD simulations from Ref. \cite{RSRB21}, 
%experimental values of G{\"u}nther and Radtke~\cite{Guenther}, 
and Lenard-Balescu results (QLB, Ronald) of Desjarlais {\it et al.} \cite{Desjarlais} and Karakhtanov~\cite{Karachtanov16}.  
$\rho^{\rm Spitzer}_1=0.846$ and $\rho^{\rm Lorentz}_1=0.492$ are defined in the text.
The green line represents a linear extrapolation of the converged DFT-MD results.
Data are given in the Supplemental material of  \cite{RSRB21}.\label{fig:3}}
\end{figure}

Numerous  studies have been performed to calculate the electrical conductivity $\sigma(n,T)$ of  Hydrogen plasma in a wide range of parameters, a recent review can be found in  Ref. \cite{Starrett20}. A comparative study~\cite{Grabowski} was also recently published that considered different approaches and showed large differences in the calculated conductivities. 
Analytical calculations in the framework of generalized linear response theory were performed for simple systems such as the Hydrogen plasma. 
For more complex plasmas, the DFT-MD approach \cite{Desjarlais02,Desjarlais,Redmer,French2017} was elaborated to evaluate the Kubo-Greenwood formula.
However, as discussed in \cite{Heidi}, electron-electron collisions are not correctly described in this approach.
In a recent study \cite{RSRB21}, the low-density limit of the electrical conductivity $\sigma (n,T)$ of Hydrogen as the simplest ionic plasma is presented as a function of temperature $T$ and particle density $n$ in terms of a virial expansion of resistivity. The  non-consideration of the contribution of electron-electron collisions in other transport coefficients such as thermopower and thermal conductivity has also been discussed recently \cite{Desjarlais,French22}. 

The virial expansion of the dimensionless resistivity $\rho^*$, Eq. (\ref{eq:5}), contains the logarithmic term $\ln(1/n)$.
 To make its argument dimensionless we use the Born parameter, see Ref. \cite{RSRB21}, 
 \begin{equation}
\frac{\Theta}{\Gamma}=\frac{T_{\rm Ry}^2}{n_{\rm Bohr}} (96 \pi^5)^{-1/3}\,,
\label{GamT}
\end{equation}
where the temperature is measured in Rydberg units, $T_{\rm Ry}=2 T_{\rm Ha}=k_BT/13.6~\textrm{eV}$.
 As discussed in Sec. \ref{sec3} in connection with the logarithmic term, we use a modified virial expansion
and rewrite  (\ref{eq:5})
\begin{eqnarray}
% \rho^*(\Gamma,\Theta)= \tilde\rho_1 (\Gamma^2\Theta) \ln \left(\frac{\Theta}{\Gamma}\right) + \tilde\rho_2 (\Gamma^2\Theta) + \dots 
 \rho^*(n,T)= \tilde\rho_1 (T) \ln \left(\frac{\Theta}{\Gamma}\right) + \tilde\rho_2 (T) + \dots \,.
 \label{eq:vir}
\end{eqnarray}
 The modified virial coefficients $\tilde\rho_i$ are related to $\rho_i$ replacing in Eq.~(\ref{eq:vir}) the variables $\Theta,\Gamma$ by $n,T$ according to
Eq.~(\ref{GamT}).
Comparing with Eq.~(\ref{eq:5}),  $\tilde\rho_1=\rho_1$ is obtained and 
\begin{equation} 
 \tilde\rho_2=\rho_2 + \rho_1 \ln[ (96 \pi^5)^{1/3} /T_\textrm{Ry}^2] \,.
\end{equation}

A highlight of plasma transport theory is that the exact value of the first virial coefficient for Coulomb systems is known from the seminal paper of Spitzer and H{\"a}rm~\cite{Spitzer53},
\begin{equation}
\label{eq:Spitzer}
 \rho_1 = \tilde\rho_1 =\rho^{\rm Spitzer}_1 = 0.846024,
\end{equation}
which does not depend on $T$. 
Note that Eq.~(\ref{eq:Spitzer}) accounts for the contribution of the electron-electron ($e-e$) interaction. In contrast, for the Lorentz plasma model where the $e-e$ collisions are neglected so that only the electron-ion interaction is considered, the first virial coefficient is \cite{Roep88}
\begin{equation}
\rho_1^{\rm Lorentz} =\frac{1}{16} (2\pi^3)^{1/2}=0.492126\,.
 \label{Lorentz}
\end{equation}
Although $e-e$ collisions do not contribute to a change of the total momentum of the electrons due to conservation of momentum, the distribution in momentum space is changed by $e - e$ collisions ("reshaping"), and higher moments of the electron distribution are not conserved by $e - e$ collisions. The indirect influence of $e-e$ collisions on the dc conductivity becomes clear in generalized linear response theory where these higher moments are considered, see~\cite{Roep88,Redmer97}.

No exact value is known for the second virial coefficient $\rho_2(T)$ or $\tilde\rho_2(T)$. 
It depends on the treatment of the many-body effects, in particular on the screening of the Coulomb potential. 
In a quantum statistical approach, the static (Debye) screening by electrons and ions
should be replaced by dynamical screening. 
For the Hydrogen plasma considered here, the Born approximation for the collision integral at high temperatures $T_\textrm{Ry}\gg 1$ is justified. Consideration of screening in the random phase approximation (RPA), 
leads to the quantum Lenard-Balescu (QLB) expression. 
Thus, at very high temperatures, where the dynamically screened Born approximation becomes valid, we obtain the QLB result, 
 see~\cite{Desjarlais,Karachtanov16},
\begin{equation}
%\lim_{T \to \infty}\rho^*=\rho_{\rm LB}^*=0.846024 \ln (\Theta/\Gamma) +0.491709
 \lim_{T \to \infty}\tilde\rho_2(T) = \tilde \rho_2^{\rm QLB} = 0.4917~.
 \label{virLB}
\end{equation}

As $T$ decreases, strong binary collisions (represented by ladder diagrams) become important and must be treated in the calculation of the second virial coefficient $\tilde \rho_2(T)$ beyond the Born approximation. According to Spitzer and H{\"a}rm~\cite{Spitzer53}, 
the classical treatment of strong collisions with a statically screened potential gives for $\rho^*=1/\sigma^*$ the result
\begin{equation}
\rho_{\rm Sp}^*=0.846 \ln \left[\frac{3}{2} \Gamma^{-3} \right]\,.
\end{equation}
Interpolation formulas have been proposed that link the high-temperature limit $\tilde\rho_2^{\rm QLB}$ with the low-temperature 
Spitzer limit \cite{Esser03,GDW,WDW,RR89,RRMK89,Redmer97,Roep88,EssRoep98}. 
Based on a T-matrix calculation in quasiclassical (Wentzel-Kramers-Brillouin, WKB) approximation~\cite{Esser03,RRT89}, 
the expression ($T_{\rm eV}=k_BT/$eV)
\begin{equation}
 \tilde\rho_2(T_{\rm eV}) \approx 0.4917 + 0.846 \ln\left[ 
 \frac{1 + 8.492/T_{\rm eV}}{1 + 25.83/T_{\rm eV} + 167.2/T_{\rm eV}^2} \right]
 \label{WKB}
\end{equation}
is a simple interpolation that combines the QLB result with the Spitzer limit in WKB approximation.
However, the exact analytical form of the temperature dependence of the second virial coefficient $\tilde\rho_2(T)$ remains an open problem. 

Thus, the available exact results for the virial expansion (\ref{eq:vir}) 
of the inverse conductivity of fully ionized 
Hydrogen plasma are:\\ 
(i) the value of the first virial coefficient is $\tilde\rho_1 = 0.846$; \\
(ii) the second virial coefficient has the high-temperature limit $\lim_{T \to \infty}\tilde\rho_2(T) = 0.4917$; \\
(iii) the second virial coefficient is temperature dependent, an approximation  is given by Eq.~(\ref{WKB}).

To extract the first and second virial coefficient from calculated or measured dc conductivities, we plot the expression 
\begin{equation}
 \tilde \rho(x,T) = \frac{\rho^*}{\ln (\Theta/\Gamma)}=\frac{32405.4}{\sigma (n,T) (\Omega {\rm m})}  T_{\rm eV}^{3/2} 
 \frac{1}{\ln (\Theta/\Gamma)}
 \label{virrho}
\end{equation}
as a function of $x=1/\ln(\Theta/\Gamma)$ and $T$ in Fig.~\ref{fig:3} which is called \textit{virial plot}.
According to Eqs.~(\ref{eq:5}), (\ref{eq:vir}), the behavior of any isotherm (fixed $T$) is linear near $n \to 0$, 
\begin{equation}
\tilde \rho(x,T) = \tilde \rho_1(T) + \tilde \rho_2(T) x + \dots\,,
 \label{virrho1}
\end{equation}
 with $\tilde\rho_1(T)$ 
as the value at $x=0$ and $\tilde\rho_2(T)$ as the slope of the isotherm.  In this way, the extraction of virial coefficients becomes immediately possible.
For $x > 1/\ln(100)=0.217$, the contributions of higher order virial coefficients  
have to be taken into account \cite{RSRB21}. For fixed $T$ and low density, where $\theta \gg 1$, a classical plasma is present and the effects of degeneracy contribute to the higher order virial coefficients.

In Fig.~\ref{fig:3} two cases for the first virial coefficient $\rho_1$ on the ordinate axis are shown, see also~\cite{Roep88,RR89,Redmer97}:\\ 
(i) $\rho^{\rm Spitzer}_1$ from kinetic theory when $e-e$ collisions are taken into account,\\
(ii) when $e-e$-collisions are neglected, $\rho_1^{\rm Lorentz}$ is obtained 
for the Lorentz plasma model.\\
Moreover, the second virial coefficient $\tilde\rho_2^{\rm QLB}$ of the Lenard-Balescu approximation. 
(\ref{virLB}) is shown, which is correct in the high temperature limit.
The QLB calculations of Desjarlais {\it et al.} \cite{Desjarlais} are shown in Fig. \ref{fig:3}.
The $e-e$ collisions are taken into account, yielding the same asymptote ($x \to 0$) as in Karakhtanov~\cite{Karachtanov16}. With increasing 
$x=1/\ln(\Theta/\Gamma)$ small deviations from linear behavior are observed. When isotherms are presented, this deviation indicates the contribution of higher virial coefficients.

Virial plots are presented in \cite{RSRB21} to investigate two problems: Which of the various approaches that give us analytical expressions for the electrical conductivity of Hydrogen plasmas are accurate in the low density limit? 
The virial expansion of the inverse conductivity serves as an exact benchmark for theoretical approaches, so that the accuracy and consistency of semi-empirical results for conductivity, such as those collected in Ref.~\cite{Grabowski}, can be checked.
A more fundamental problem is whether numerical results from molecular dynamics simulations based on density functional theory (DFT-MD) correctly contain the contribution of electron-electron collisions.
The virial plot confirms the position that DFT-MD simulations in the low-density limit describe a Lorentz plasma with only electron-ion collisions, the contribution of electron-electron collisions to $\rho_1$ is  missing \cite{RSRB21}.

Here we discuss some details of the virial expansion for the inverse conductivity and the corresponding virial plots, see Fig.  \ref{fig:3}.
DFT-MD simulations are given in Ref. \cite{RSRB21}, see the tables of data in the supplementary material.
These data have sufficiently high accuracy, as can be seen from the small deviations from the fit line in Fig. \ref{fig:3}.
In addition to the precise solution of the Kubo-Greenwood formula, this is achieved by good control of convergence with increasing particle number, as shown by comparison of calculations with different numbers of particles.
The number of particles must be sufficiently large to ensure convergence. In the parameter range considered in the figure, about 100 particles in the box are necessary to achieve convergence.
Further calculations with 216 electrons 
were not possible due to limited computer capacity. For $T=150$~eV, even 125 electrons exceed the currently available 
computer capacity.
This point was also discussed in a recent work \cite{French22}, where earlier calculations \cite{Desjarlais} were improved to achieve convergence. 
Another problem is the determination of the value of the dc conductivity $\sigma(0)$ from the calculation of the optical conductivity $\sigma(\omega)$ at finite frequencies. 
Because of the discretisation in a finite box, the energy eigenvalues have a minimum spacing and the energy-conserving $\delta$ function must be smeared by a parameter $\epsilon$ to allow for transitions, see also section \ref{sec2} above. 
To reach the limit $\omega \to 0$, an extrapolation is performed according to the Drude formula (\ref{eq:Drude}). This was discussed also in Ref. \cite{French22}. Instead, one can use the dynamic collision frequency to perform this extrapolation.

The results shown in Fig. \ref{fig:3} allow the extraction of virial coefficients $\rho_1(T), \tilde{\rho}_2(T)$. Compared to other approaches, including interpolation formulas, see \cite{RSRB21}, as well as the QLB calculation, we assume that we are in the linear region of the virial curve. Deviations from linearity can be observed for QLB already at $x=0.2$, since the density is high (40~g/cm$^3$). For DFT-MD simulations with density about 2 g/cm$^3$, the deviation from linearity for the last point is observed at $x\approx 1$. 

As pointed out in \cite{RSRB21}, the extrapolated value of $\rho_1$ in the virial plot at $x=0$ points to the Lorentz value (\ref{Lorentz}) but misses the Spitzer value (\ref{eq:Spitzer}). This means that electron-electron collisions are not considered in the DFT-MD calculations for the electrical conductivity. Also of interest is the value of $\tilde{\rho}_2(T)$ given by the slope in the virial plot near $x=0$.
Fitting it to the data gives a slope of 0.9886 for the DFT-MD calculations. This is about twice the slope $\tilde \rho_2^{\rm QLB}$ given above. From analytical approaches, it appears that the slope is determined by various effects such as dynamical screening and strong collisions. In the limiting case of high temperatures, the Born approximation should be possible, but the Coulomb potential must be replaced by a screened potential. Static screening of the proton scatterer with both electrons and protons would lead to the following result ($C=0.57721\dots$ is Euler's constant).
\begin{equation}
\lim_{T \to \infty} \tilde{\rho}_2(T)=\frac{\pi^{3/2}}{24 \sqrt{2}}\left[\frac{11}{2}-3 C+\ln\left(\frac{3}{2} \pi^2\right)\right] =1.06036
\end{equation}
which is close to the observed slope of the DFT-MD simulations. However, it remains unclear to what extent the screening is included in the simulations. We assume that the ionic structure factor, which is the ionic contribution to the screening, is well described, and that the electron screening is also captured by the exchange-correlation functional. However, we need to consider dynamical screening, a problem that has been discussed in previous work \cite{RR89}
 on virial expansion.

\begin{figure}[t]
\centerline{\includegraphics[width=0.7 \textwidth]{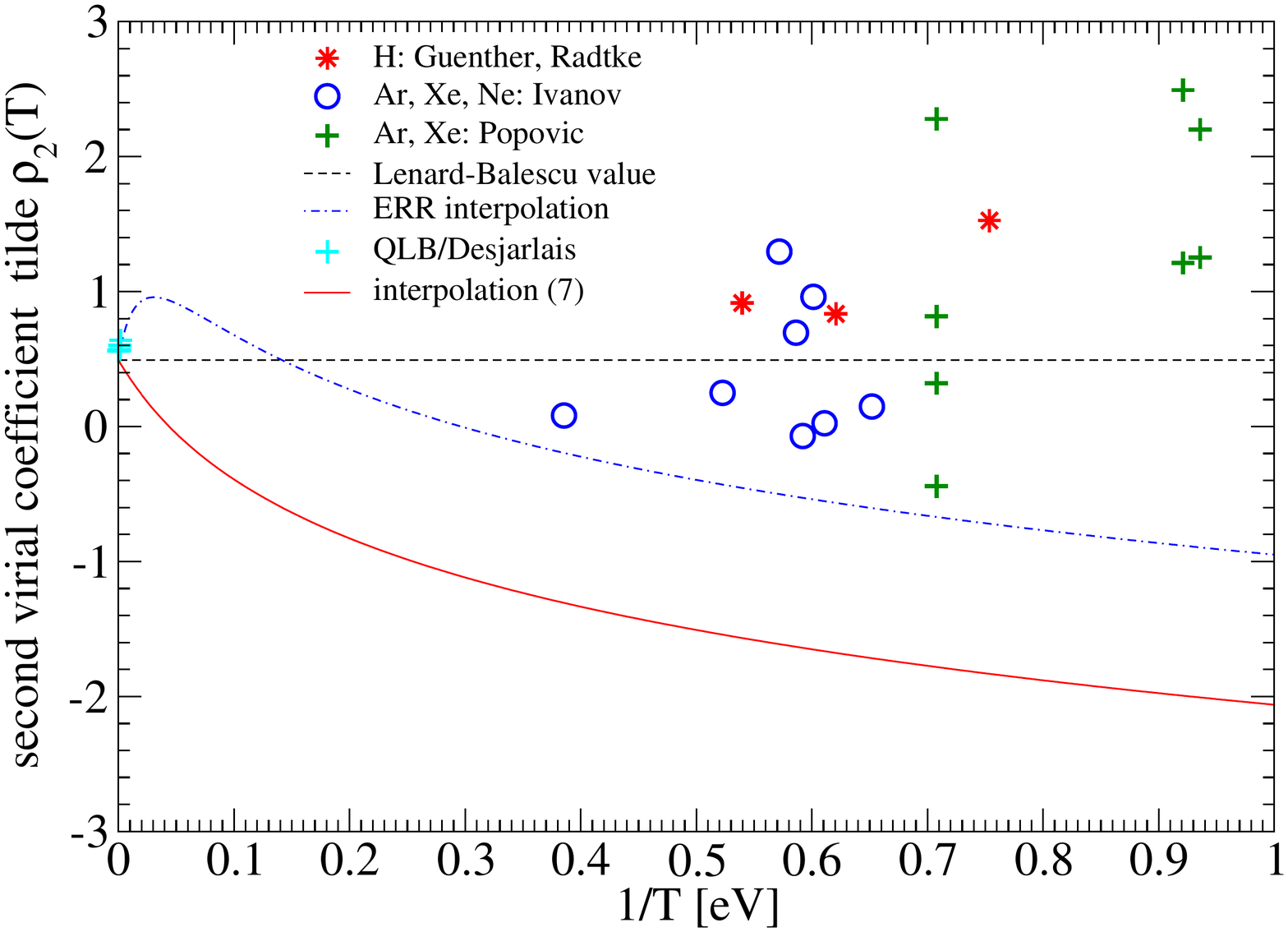}}
\caption{Second virial coefficients $\tilde\rho_2(T)$ and $\tilde\rho_2^{\rm eff}(n,T)$ for the dc conductivity of Hydrogen plasmas. Analytical interpolation formulas (\ref{WKB}) and Ref. \cite{Esser03} are compared with experiments of G{\"u}nther and Radtke~\cite{Guenther} for H plasmas as well as of Ivanov {\it et al.}~\cite{Ivanov} and Popovic {\it et al.}~\cite{Popovic} for rare gas plasmas. The black dashed line corresponds to the high temperature limit that is given by the quantum Lenard-Balescu value. The broken blue line is the interpolation formula of Ref. \cite{Esser03}, the red full line represents the interpolation formula (\ref{WKB}) for the second virial coefficient.
\label{fig:4}}
\end{figure}

We return to the long-debated question of whether or not $e-e$ collisions are accounted for in the DFT-MD formalism. For example, it was pointed out in Ref.~\cite{Heidi} that a mean-field approach is not able to describe two-particle correlations, in particular $e-e$ collisions.  
However, to some approximation, the $e-e$ interaction is accounted for by the exchange-correlation energy. 
DFT-MD simulations, which are mean-field theories that account for the $e-e$ interaction only through the exchange-correlation part of the energy density, 
cannot account for the effect of $e-e$ collisions on the conductivity, so that $\rho_1(T)$ corresponds to the Lorentz plasma, but $\tilde\rho_2(T)$ is determined by screening. 
The question arises to what extent dynamical screening, as implemented in the QLB calculations, is also described by the exchange-correlation part of the energy density functional.
We would like to mention that in the case of thermal conductivity it has been shown that 
the contribution of $e-e$ collisions is not taken into account in DFT-MD simulations \cite{Desjarlais,Starrett20,French22}
and yields an additional term. Other approaches such as generalized linear response theory may be considered to indicate appropriate approaches.

Our analysis has shown that the simulation results with virial evolution 
are extrapolated to the low-density region, where DFT-MD simulations are no longer feasible. The current simulations, while computationally expensive, are still not very close to $x=0$, so extrapolation to the $x=0$ limit is not very accurate. Better data for DFT-MD simulations would be of interest to confirm our results. Conversely, the benchmark capability of virial expansion discussed in this work can also serve as a criterion to verify the accuracy of numerical approaches such as DFT-MD simulations to evaluate conductivity.

\begin{table}[htp]
\caption{Experimental data for the electrical conductivity. 
 G{\"u}nther and Radtke: H~\cite{Guenther}; 
 Ivanov {\it et al.}: Ar, Xe, Ne~\cite{Ivanov}; 
 Popovic {\it et al.}: Ar, Xe~\cite{Popovic}.}
\begin{center}
\begin{tabular}{|c|c|c|c|c|c|c|c|c|c|c|}
\hline
Plasma &$\hat n_e \times 10^{25}$& $n\times 10^{-6}$ & $T \times 10^{4}$& $T$& $\Gamma$ & $\Theta$ & $1/\ln(\Theta/\Gamma)$ & $\sigma  \times 10^3 $ &$\tilde \rho (x,T)$  & $\tilde \rho_2^{\rm eff}$\\
&$[{\rm m}^{3}]$& [g/cm$^3$]  & [K]&  [eV] &  & &  &$[(\Omega {\rm m})^{-1}]$ &  & \\
\hline
H&$0.1 $	& $1.67262$ 	& $1.54$ & 1.32706  & 0.174914 & 363.932 & 0.130883& 6.2 & 1.04579 & 1.52647 \\
H&$0.15$	& $2.50893  $  	& $1.87 $ & 1.61143  & 0.164892 &337.249 & 0.131177& 9.1& 0.955544 & 0.835097\\
H&$0.25$	& $4.18155 $ 	& $2.15$ & 1.85271 &0.170041 & 275.832 & 0.13529& 11.4 & 0.969821 &  0.915228\\
Ar&$2.8$	& $46.8334 $   & $2.22 $ &1.91303  &0.36845 & 56.8959 & 0.198426& 19.0 &0.895459 & 0.249255 \\
Ar&$5.5$	& $91.9942 $   & $2.03$ & 1.74931  &0.504626 &33.1707 & 0.238914& 15.5 & 1.15565 & 1.29607 \\
Ar&$8.1$	& $135.482$   	& $1.93 $ & 1.66313  &0.603878 & 24.3632 & 0.270456& 17.0 & 1.10575 & 0.960407 \\
Ar&$14$		& $234.167 $   & $1.9 $ & 1.63728 &0.736152 & 16.6533 & 0.320623& 25.5  & 0.853604 & 0.0237179 \\
Ar&$17$		& $284.346 $   & $1.78$ & 1.53387  &0.838316 & 13.7074 &0.357872& 24.5  & 0.899216 & 0.148701 \\
Xe&$25$		& $418.155 $   & $3.01 $ & 2.5938  &0.563757 & 17.9242 & 0.289077& 45 & 0.869607 &  0.081664\\
Ne&$1.1$	& $18.3988 $   & $1.98 $ & 1.70622  &0.302559 & 94.6027 & 0.174059& 13  & 0.966995 & 0.695135\\
Ne&$1.9$	& $31.7798$   	& $1.96 $ & 1.68899  &0.366725 & 65.0509 & 0.193113& 16.5  & 0.832499 & -0.0699113 \\
air&$0.13$	& $2.17441 $   	& $1.1 $ & 0.9479  &0.26726 & 218.238 & 0.14914& 6& 0.743367 & -0.688167 \\
Ar&$0.06$	& $1.00357 $   	& $1.64$ & 1.41323  &0.138532 & 544.807 &0.120816 &8.3 & 0.792469 & -0.443077 \\
Ar&$0.1$	& $1.67262$   	& $1.64  $ & 1.41323 &0.164248 & 387.564 & 0.128762& 7.9  & 0.887358 & 0.321199 \\
Ar&$0.13$	& $2.17441 $   & $1.64 $ & 1.41323  &0.17926 & 325.373 & 0.133264& 7.6  & 0.954636 & 0.815191 \\
Ar&$0.15$	& $2.50893 $   & $1.64 $ & 1.41323 &0.188017& 295.767& 0.135855&  6.4  & 1.15567 &  2.27941\\
Xe&$0.06$	& $1.00357$   	& $1.24 $ & 1.06854  &0.18322 & 411.928 & 0.129569& 4.6\ & 1.0082 & 1.25185 \\
Xe&$0.12$	& $2.00715 $   & $1.24$ & 1.06854 &0.18322 & 411.928& 0.13529& 4.1  & 1.0082 &2.20078  \\
Xe&$0.07$	& $1.17083$  	& $1.26$ & 1.08578&0.189819&377.693&0.131652& 4.8 & 1.00558 & 1.21211\\
Xe&$0.14$	& $2.34167$ 	& $1.26$ & 1.08578  &0.239157 &237.931 & 0.144873& 4.4  & 1.20715 & 2.49289 \\
\hline
\end{tabular}
\end{center}
\label{tab:7}
\end{table}%

Another application of the virial plot is experiments to measure electrical conductivity.
Assuming that the value $0.846024$, Eq. (\ref{eq:Spitzer}), for $\rho_1$ is exact, an effective second virial coefficient 
\begin{equation}
 \tilde\rho^{\rm eff}_2(n,T) = \frac{32405.4}{\sigma (n,T) [\Omega {\rm m}]} \left(\frac{T}{{\rm eV}}\right)^{3/2} 
 - 0.846024 \,\, \ln\left(\frac{\Theta}{\Gamma}\right)
 \label{rho2eff}
\end{equation}
has been introduced which gives the second virial coefficient in the low-density limit,  
$\lim_{n \to 0} \tilde\rho^{\rm eff}_2(n,T)= \tilde\rho_2(T)$.
A dependence of  $\tilde\rho_2^{\rm eff}(x,T)$ on density shows that higher orders of the virial expansion are relevant.
We anticipate that at very high $T$, i.e., $1/T\to 0$, the Lenard-Balescu value is approximated. 
The deviations at increasing $1/T$, shown in the interpolation formula and the DFT-MD simulations, indicate 
that already below temperatures of the order of 100~eV, the effect of strong collisions beyond the Born 
approximation should be taken into account.

Ultimately, the virial expansion (\ref{eq:vir}) must be verified experimentally, but 
accurate data for the conductivity of Hydrogen plasma in the low-density limit and/or 
at high temperatures are scarce. Accurate conductivity data for dense Hydrogen plasma 
were derived by G{\"u}nther and Radtke~\cite{Guenther}.
%, which were also shown in the virial 
%plot, Fig. \ref{fig:3}. 
They are close to the benchmark data of the virial expansion.
It should be noted that there are systematic errors associated with the analysis of such experiments. 
For example, the appearance of bound states requires a realistic treatment of the plasma 
composition and the influence of neutrals on electron mobility. Alternatively, 
conductivity measurements in highly compressed noble gas plasmas were carried out by 
Ivanov {\it et al.}~\cite{Ivanov} and Popovic  {\it et al.}~\cite{Popovic,Esser03}, 
but the interaction of the electrons with the ions deviates from the pure Coulomb 
potential due to the core of bound electrons. The corresponding virial plot is close 
to the data of Hydrogen plasma, see \cite{RSRB21}, but requires a more detailed 
discussion on the role of bound electrons. 

It should also be mentioned that the densities are quite high, 
and extrapolation to zero density must be performed to obtain the second virial coefficient.
This tendency can be seen in Fig.~\ref{fig:4}, especially for the experiments with Ar, Xe~\cite{Popovic},
where low-density data point to $\tilde\rho_2(T)$. 

Quantum statistical methods provide accurate values for the lowest virial coefficients, which serve as benchmarks for analytical approaches to electrical conductivity as well as for numerical results from molecular dynamics simulations based on density functional theory (DFT-MD) or path integral Monte Carlo (PIMC) simulations. 
While these simulations are well suited to compute $\sigma (n,T)$ in a wide range of densities and temperatures, especially for the warm dense matter region, they become computationally expensive in the low density limit, and virial expansions can be used to compensate for this drawback. Interpolation formulas that take both approaches into account would be very useful for calculating the conductivity of plasmas. 

To obtain the correct values for the thermoelectric transport coefficients of Hydrogen plasma in the low-density limit, where the inclusion of $e-e$ collisions is essential,
different solutions can be considered. 
PIMC simulations, as successfully performed for the uniform electron gas \cite{TD}, should also be performed for the two-component plasmas.  
First steps of this ambitious project are recently in progress \cite{Bonitz20,Boehme22}. 
The study of such PIMC calculations with the virial plot would be of great interest.
From generalized linear response theory, we also learn that higher order correlation functions, such as force-force correlation functions associated with the dynamic collision frequency, 
may be a suitable approach to include the contribution of $e-e$ collisions in the transport coefficients \cite{Roep88,RR89,Redmer97}.

\section{Conclusions}\label{sec7}

We have from quantum statistics exact expressions for  thermodynamic and transport properties of plasmas by equilibrium correlation functions,
but the evaluation is a complex problem in many-particle physics. 
Numerical simulations are becoming more accurate as computer capacity increases.
However, they need to be controlled with respect to their limits such as size effects, but also fundamental problems such as the correct description of electron-electron collisions in the context of DFT or the sign problem in PIMC.
It is expected that PIMC simulations will provide an adequate description of electron-electron interactions, but they are currently unable to solve complex plasmas such as multiply charged ions in the low-temperature range.

The comparison of analytical results for the virial expansion of thermodynamic properties with  PIMC calculations for the uniform electron gas has been performed. 
In particular, we show that high-precission PIMC simulations confirm the correct form of the virial expansion, which has been debated recently. 
It seems to be possible to give also numerical values for higher virial coefficients, in particular the interesting $n^{5/2}$ coefficient. 
These values can be considered as exact results in plasma physics. Numerical values for higher virial coefficients would also be of great interest for transport properties.

Analytical theory gives us exact results in limiting cases. This can be used to obtain results for parameter ranges where numerical simulations are not efficient, e.g. in the low density range. Virial expansions are used to control theories and numerical simulations. They are of interest to construct interpolation formulas.

It was indicated that the evaluation of the Kubo-Greenwood formula using DFT-MS simulations does not take into account the effects of electron-electron scattering and cannot reproduce the low-density limit of the electrical conductivity of Hydrogen plasmas. 
Similar results were recently reported by French {\it et al.}  \cite{French22} for other thermoelectric transport coefficients. It would be of interest to perform PIMC simulations that can accurately describe electron-electron collisions.

The theory of virial expansion must be extended if the formation of bound states is of importance, i.e. for $T/T_{\rm Ha} \le 1$, see appendix.  New approaches are needed.
The approach described here is also applicable to other correlation functions such as the dynamic structure factor or to other transport properties such as thermal conductivity, thermopower, viscosity, and diffusion coefficients. 
Also of interest is the extension of virial expansion to elements other than Hydrogen, where different ions can be formed and the electron-ion interaction is no longer purely Coulombic. 

\section*{Acknowledgments}
Thanks to M. Sch\"orner, R. Redmer, M. Bethkenhagen, M. French, H. Reinholz, T. Dornheim, J. Vorberger, Z. Moldabekov, and W.-D. Kraeft for collaboration and discussions.
This work was supported by  the German Research Foundation
(DFG), Grant \# RO 905/37-1 AOBJ 655625.

\subsection*{Author contributions}

This is an author contribution text. It is based on a contribution to the SCCS22 conference.

\subsection*{Financial disclosure}

None reported.

\subsection*{Conflict of interest}

The author declares no potential conflict of interests.

\appendix
\section{Bound state formation}\label{sec6}

A special problem of plasmas is the formation of bound states (atoms, charged ions: clusters with a certain number of elementary particles, i.e., nuclei and electrons) which can dominate the properties in the low-density and low-temperature region. 
A simple approach is the chemical picture \cite{KKER}, where the bound states are considered as new constituents. 
The interaction between the different constituents is neglected except for reactive collisions. Thus, a chemical equilibrium is achieved in which the composition of the plasma is described by the law of mass action. For a systematic approach including bound state formation see Refs. \cite{AB10,AB12} and references given there.
We will not present here an exhaustive discussion of the chemical picture, but only discuss some aspects in the context of our work. For a recent review, see \cite{ERR,ER22}, where further references can be found. 

Within the chemical picture, several issues arise that need to be discussed in order to improve this simple approximation, using the concept of virial expansions.\\
(i) In addition to the ground state, excited states ($s$) with excitation energy $E_{\alpha, s}$ can occur, which can also be treated as new species. It is more convenient to introduce the intrinsic partition function of the cluster $\alpha$, which is summed over all excited bound states by the statistical factor $\exp[-\beta E_{\alpha, s}]$.\\
(ii) In addition to bound states, there are also scattering states that must be included in the calculation of virial coefficients. This leads to the Beth-Uhlenbeck formula, in which the scattering phase shifts appear.
Sometimes resonances can appear in the spectrum of excited states. In the resonance gas approximation, the intrinsic partition function is improved by extending the summation over all excitations $s$ to the resonances in the continuum. Moreover, the contribution of scattering phase shifts should be included.\\
iii) We arrive at higher virial coefficients and need to include density effects. In the framework of a quasiparticle approach,
the intrinsic partition functions are calculated with shifted energies due to screening, mean-field effects, Pauli blocking and other effects. 
%The composition is calculated as before.

As example, let us consider the H plasma and give the intrinsic partition function in the simplest approximation 
\begin{equation}
\label{intrH}
z_{\rm H} = \sum_s 2 s^2 e^{-E_{{\rm H},s}/k_BT}
\end{equation}
with the known energy levels $E_{{\rm H},s}= -E_{\rm Ha}/(2 s^2)$ ($E_{\rm Ha} =27.2$ eV is the Hartree energy). The factor $2 s^2$ denotes the degeneracy of the excitation $s$ including the spin factor. As specific for the Coulomb interaction, we have infinitely many bound states near the continuum edge for $s \to \infty$. Expression (\ref{intrH}) is not applicable because it is divergent. A convergent expression is the Planck-Brillouin-Larkin partition function, see \cite{KKER},
\begin{equation}
z_{\rm H} = \sum_s 2 s^2 \left[ e^{-E_{{\rm H},s}/k_BT}-1+\frac{E_{{\rm H},s}}{k_BT}\right].
\end{equation}
The subtraction of 1 is explained as follows: We need to include the contribution of the scattering states which compensate for the most divergent term of the contribution of the bound states.  
For the short-range interaction, this has been discussed in detail, and generalized phase shifts have been introduced to avoid separating the bound and scattering parts of the intrinsic partition function \cite{nucl,R15}.

More complex is the explanation of the subtraction of $E_{\rm Ha}/(2 s^2k_BT)$. Because of the long-range character of the Coulomb interaction, phase shifts cannot be defined in the usual form, 
and the contribution of the scattering states is not well defined when scattering phase shifts are used. 
This fundamental problem of the Coulomb interaction is solved introducing the concept of screening. 
In the framework of a quantum statistical approach, we have to perform the partial sum of so-called ring diagrams and introduce quasiparticles. We must, however, avoid double counting. This has already been discussed in detail for the Hartree-Fock approximation \cite{SRS,RBBKTW13}. Of interest is the generalization to partially ionized plasmas with multiply charged ions \cite{ER22}.

A systematic approach arises from consideration of the spectral function. 
We can identify a quasiparticle contribution and perform a cluster decomposition of the self-energy. 
For the cluster decomposition of the self-energy, we can introduce different channels. 
To avoid double counting, diagrams used for the single-particle self-energy must be subtracted from the ladder sums defining the cluster states.

A related problem is the definition of the ionization degree in dense plasmas, since the separation of the bound state contribution from the intrinsic partition function is arbitrary,  see\cite{Lin17,GR19,Bethkenhagen20} and references given there. 
A possible solution would be the definition of the single-quasiparticle contribution which is extracted from the spectral function.
Thus it can be performed by considering the compressibility or the dynamical conductivity.

The inclusion of bound states and the corresponding generalization of the chemical picture, involving quasiparticle concepts for the free and bound states, is a difficult problem in plasma theory. 
Of course, at fixed temperature there is always a low-density limit at which bound states are dissolved (because of entropy) but this regime can be very limited, for instance it is not applicable to gases under normal conditions. 
A realistic description is often based on the chemical picture where bound states are considered, i.e. for temperatures below the binding energies.
A generalized quasiparticle approach is well defined at low densities, but has to be generalized considering the spectral function (\ref{spectral}) if densities are increasing.
The formation of bound states is not only important for the thermodynamic properties, as discussed above for the second virial coefficient of the Hydrogen plasma. 
It also determines the transport properties, and the consideration of bound states as additional scatterers remains a complex problem if we want to go beyond the simple chemical picture.

%%%%%%%%%%%%%%%%%%%%%%%%%%%

\end{document}